\documentclass[fleqn,usenatbib,useAMS]{mnras}
\pdfoutput=1

\usepackage{graphicx}	
\usepackage{amsmath}	
\usepackage{amssymb}	
\usepackage{multicol}        
\usepackage{bm}		
\usepackage{pdflscape}	

\usepackage[T1]{fontenc}
\usepackage{ae,aecompl}

\usepackage{newtxtext,newtxmath}
\usepackage{threeparttable}

\title[Ca-rich SNe \& sGRBs ofsets distribution]{No velocity-kicks are required to explain large-distance offsets of Ca-rich supernovae and short-GRBs}

\author[Perets \& Benamini]{Hagai B. Perets$^{1}$
 and Paz Beniamini$^{2,3}$\\
$^{1}$ Physics Department, Technion --- Israel Institute of Technology, Haifa 3200003, Israel\\
$^{2}$Theoretical Astrophysics, Walter Burke Institute for Theoretical Physics, Mail Code
350-17, Caltech, Pasadena, CA 91125, USA \\
$^{3}$Astrophysics Research Center of the Open University (ARCO), The Open University of Israel, P.O Box 808, Ra’anana 43537, Israel\\
}

\begin{document}
\label{firstpage}
\pagerange{\pageref{firstpage}--\pageref{lastpage}}
\maketitle

\begin{abstract}
The environments of explosive transients link their progenitors to the underlying stellar population, providing critical clues for their origins. However, some Ca-rich supernovae (SNe) and short gamma ray burst (sGRBs) appear to be located at remote locations, far from the stellar population of their host galaxy, challenging our understanding of their origin and/or physical evolution. These findings instigated models suggesting that either large velocity kicks were imparted to their progenitors, allowing them to propagate to large distances and attain their remote locations; or that they formed in dense globular clusters residing in the halos. Here we show that instead, the large spatial-offsets of these transients are naturally explained by the observations of highly extended underlying stellar populations in (mostly early type) galaxy halos, typically missed since they can only be identified through ultra-deep/stacked images. Consequently, {\emph no} large velocity kicks, nor halo globular cluster environments are required in order to explain the origin of these transients.  These findings support thermonuclear explosions on white-dwarfs, for the origins of Ca-rich SNe progenitors, and no/small-natal-kicks given to sGRB progenitors.  Since early-type galaxies contain older stellar populations, transient arising from older stellar populations would have larger fractions of early-type galaxy hosts, and consequently larger fractions of transients at large offsets. This is verified by our results for sGRBs and Ca-rich SNe showing different offset distributions in early vs. late-type galaxies. Furthermore, once divided between early and late type galaxies, the offsets' distributions of the different transients are consistent with each other. Finally, we point out that studies of other transients' offset distribution (e.g. Ia-SNe or FRBs) should similarly consider the host galaxy-type.
\end{abstract}



\section{Introduction}
The host galaxies and the positions of explosive transients inside their hosts hold critical information about their progenitors. The progenitors of supernovae (SNe) and gamma-ray bursts (GRBs) can be potentially linked to the underlying stellar populations in their close environments. 

 For example, the star formation history of the host galaxy of a given transient provides a statistical inference of the age of the progenitor of the transient. Given a statistical sample of transients of a specific type, one can even infer the delay time distribution for that type of explosive events \citep[e.g.][]{Mao+14}. Such inference has a long history, with Baade noting already early-on that type II SNe explode in early type (spiral) galaxies, while type I SNe explode both in early and late type galaxies, providing the first clue of our current understanding of type II SNe arising from core-collapse SNe, while type I SNe arising from both core-collapse (most Ib, Ic SNe) and thermonuclear SNe (Ia SNe).
The local environments of transients inside their host galaxy provide further information on the stellar populations of their progenitors, their metallicities and typical lifetimes. 

In some cases, transients were found to be located in a different environment than expected from their theorized progenitors. Those cases assisted in identifying and eventually classifying such objects as part of a new class of objects arising from different type of progenitors. Two examples, which are also the focus of this study are short-GRBs (sGRBs) and Ca-rich SNe. 
For instance, long GRBs are typically observed in young environments in late type type star-forming galaxies and close to star-forming regions in these galaxies, pointing to their likely origin from massive young stars \citep{Blo+02}. In contrast, sGRBs were found both in late and in early type galaxies, and in many of which far from any star-forming region \citep{Fong+13,Fong2017}. This was found to be consistent with the idea of sGRBs arising from the mergers of neutron stars (NSs; and confirmed through the recent multi-messenger gravitational-wave and electromagnetic counterpart GRB 170817A \citep{Abb+17}). Similarly, some faint type Ib SNe were found in late-type galaxies, and far from star-forming regions, in contrast with the expected massive progenitors for core-collapse SNe typically thought to produce type Ib SNe \citep{Per+10}. This gave rise to the identification of a new class of Ca-rich, faint type Ib SNe, suggested to arise from thermonuclear He-rich explosions of white-dwarfs \citep{Per+10}, as we further discuss below. 

The early type host galaxies and the sometimes large distances from any star-forming regions of both sGRBs and Ca-rich SNe point to old progenitors for a significant fraction of these transients. This is consistent with the suggested NS mergers and thermonuclear WD explosions which can indeed have very long delay-time since the formation of their progenitor stars. However, in many cases, not only were these transients far from any star-forming regions, but they also appeared to be located very far from their host galaxies \citep{Per+10,Kasl+12,De+20,Fong+13,Fong2017,Lun+17}. It was therefore suggested by several studies that the remote locations of such transients could be explained either by large velocity kicks given to their progenitors (\citep{Lym+14,Fol+15,Lym+16} for Ca-rich SNe;\citep{Fong+13} for sGRBs), or that the progenitors form in dense globular clusters, since many of those reside in the halos \citep{Yua+13,Sel+15,She+19}. In the former scenario, the post-kick progenitors could then propagate through their host galaxies over hundreds of Myrs attaining remote positions at the time of their explosions. In particular, the observed high velocities of pulsars point to natal kicks given to NSs upon their formation. Since sGRBs are produced through the merger of NSs, it was thought plausible that sGRBs could be found at large offsets from their host galaxies. However, as we discuss below, it is debated whether indeed sGRB progenitors receive such large kicks. 

Following a similar reasoning it was suggested that many of the progenitors of Ca-rich SNe receive some large velocity kicks, either due to interactions of their progenitors with massive black holes in galaxy nuclei leading to their ejection as hypervelocity stars, or possibly due to their progenitors containing a NS which received a large natal kick.  

Already a few years ago, we pointed out that the recent findings that the stellar halos of early type galaxies are far larger than previously thought, would give rise to transients from old stellar population at these halos \citep{Per+14}. Since the underlying stellar populations can only be seen through deep imaging and/or stacking, transients arising from the halo stellar population in such galaxies would appear to be located far from the stellar populations of their host galaxies, while, in fact, they formed in-situ from stars in the difficult-to-observe underlying population.  

Here we follow this suggestion and study the detailed distributions of Ca-rich SNe and SGRBs, while dividing them between early and late type galaxies, which posses very different underlying stellar halos.  As we discuss in the following, we find that the large offsets observed for both sGRBs and Ca-rich SNe can indeed be naturally explained without invoking any (or large) natal kicks, and their progenitors could have been formed in-situ. 

The paper is structured as follows. We first discuss the distribution of the stellar masses in early vs. late type galaxies (section \ref{sec:halos}), we then analyze the offsets distributions of short Ca-rich SNe and sGRBs (section \ref{sec:offsets}). We discuss our results and their implications for Ca-rich SNe progenitors and the natal kicks given to sGRB progenitors in section \ref{sec:discussion}, and summarize (section \ref{sec:summary}). 

\section{The distribution of stellar mass in galaxy halos}
\label{sec:halos}
Traditionally, only small fractions of the stellar mass in galaxies were directly observed at large offsets (10-100 kpcs; a region which we term here the galaxy halo) from their centers. The halo of the Milky-way galaxy, for example, contains only 1-2 percents of the stellar mass in the Galaxy.  For this reason, identification of transients at large offsets from the centers of their host galaxies was typically suggested to originate from some type of velocity kicks given to their progenitors, as such stellar progenitors were thought unlikely to exist at such large offsets (especially when such large offsets are frequent for some specific type of transients). In particular, large kick velocities could allow them to migrate to large distances from their original birth place in the inner parts of the galaxy and/or their disk components, where most of the stars were expected to exist.

However, the distribution of stellar mass in galaxy halos, far from the nucleus is difficult to measure given the low surface brightness in these regions. The direct measurement of the stellar mass in galaxy halos for any single galaxy is notoriously difficult and could significantly underestimate the stellar mass in the halo. In fact, galaxy formation simulations have suggested that many galaxy halos should in fact contain significant fractions of the stellar mass \cite[e.g][and references therein]{San+18}.

Only in the last few years did the stellar mass in the halos of galaxies were measured for different types and different masses of galaxies, over large statistical samples, allowing to infer the halos' stellar masses. Such measurements were done through the use of stacking of observations of many galaxies belonging to the same type and mass range \citep{dso+14}, or through deep imaging of single galaxies using the Hyper Suprime-Cam \citep{Hua+18}. 
These studies have shown that early type galaxies (ellipticals and S0 galaxies; more centrally concentrated galaxies) have 30-70 percents of their stellar mass in the halo (with a monotonically rising trend with galaxy mass), while late type, spiral galaxies have only 0-25 percents of their stellar mass in the halo (consistent with the case of the Milky-way). 

The stellar population in early type galaxies is old, typically at least 9-10 Gyrs old \cite[e.g.]{Gal+05}, and no star formation is typically observed in galaxy halos .
Taken together, one would expect transients arising from young stellar populations to be observed in the disks of spiral galaxies, and not in galaxy halos, and thereby generally have small (<10 kpc) offsets. In contrast, transients arising from old stellar progenitors should be observed with comparable numbers in the halo and the central parts of early-type galaxies, and mostly in the disks and central parts of late-type galaxies (since only a small fraction of the old stellar population resides in the halo).

As we discuss in what follows, these recent developments in understanding the distribution of the stellar mass in galaxies have far reaching implications for the interpretations of the offsets' distribution of transients, and its ramifications regarding the progenitors of such transients, and the physical processes involved in their evolution. In particular, observations of large offsets for some types of transients such as sGRBs and Ca-rich faint supernovae should not imply the need for some natal velocity kicks for their progenitors, but rather that such large-offset transients arise from old stellar populations, and were not likely to receive large natal kick and then migrate to their observed position. As we show below, the correlation between the type of host galaxies and the offset distribution of the transients corroborates this.

\section{Ca-rich supernovae and sGRB offsets' distributions}
\label{sec:offsets}

In the following we study the offset distribution of the two type of transients, Ca-rich SNe and sGRBs. The samples are summarized in Tables \ref{tbl:SNe-off} and \ref{tbl:GRBoff}. Their cumulative distributions, divided between early and late type galaxies are shown in Fig. \ref{fig:offsets}. Descriptions of the two samples are summarized below. 

\subsection{Ca-rich supernovae sample}
\label{sec:Ca-rich-sample}
\begin{figure}
\centering
\includegraphics[width = 0.45\textwidth]{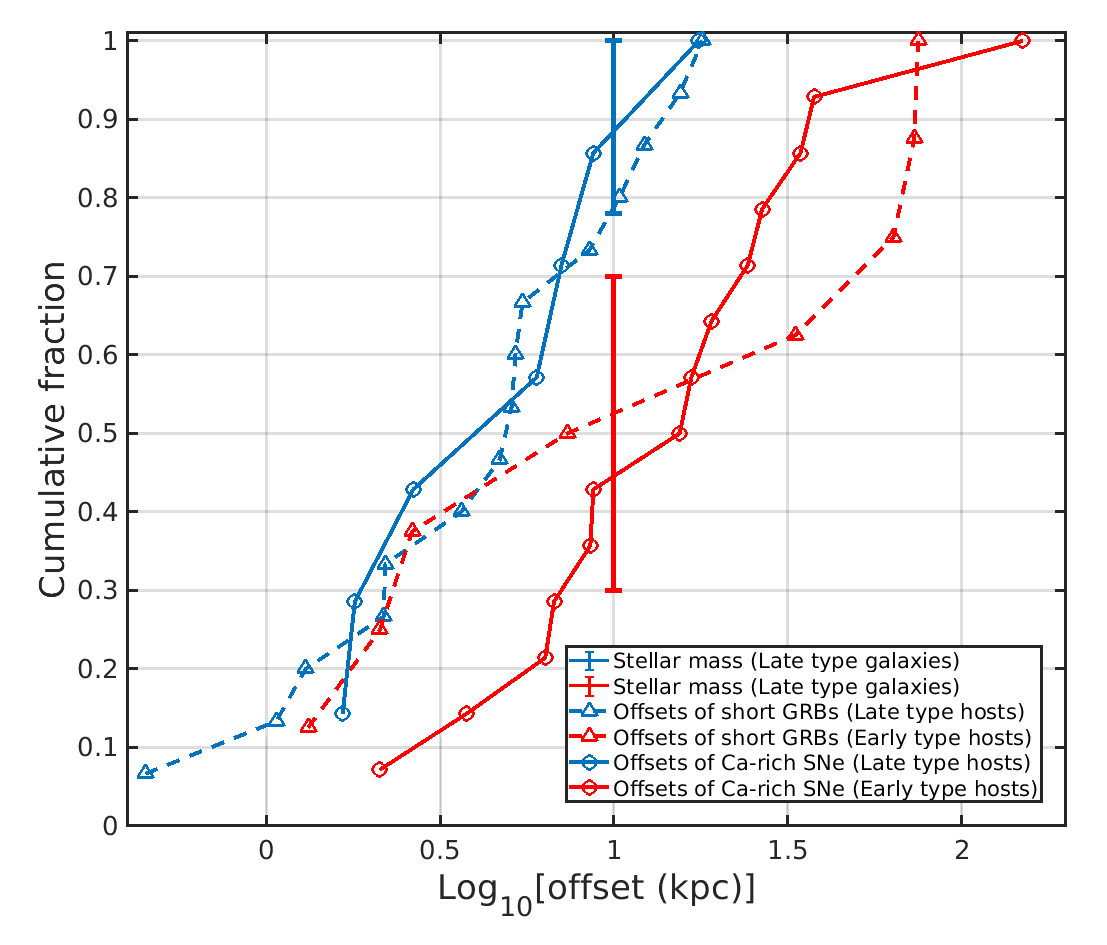}
\caption{The cumulative offsets distributions of sGRBs and Ca-rich supernovae in early and late type galaxies. The offsets distributions of both SGRBs and Ca-rich SNe, differs between the early type host galaxies and the late type host galaxies.  The offsets distribution is far more extended in the former, compared with latter,  consistent with the far more extended distribution of the halo stellar mass in early type galaxies, compared with late type galaxies, and consistent with the observed stellar masses in these types of galaxies (vertical lines, from \citealt{dso+14}). These results suggest an in-situ formation for the progenitors of these transients, without the need of any natal velocity kicks, nor the existence of globular clusters.}.
\label{fig:offsets}
\end{figure}

 Our current sample of Ca-rich SNe offsets is based on the sample from our original characterization study \citep{Per+10} based on SNe from the LOSS and CCCP surveys, and the additional Ca-rich SNe discovered by the PTF/iPTF \citep{Kasl+12,Lun+17,De+18}, PESSTO \citep{Val+14} and ZTF \citep{De+20,Wyn+20} surveys. The SNe offsets are taken from these studies, and summarized in Table \ref{tbl:SNe-off}; the cumulative offset distributions divided between early and late type galaxies can be found in Fig. \ref{fig:offsets}. The error bars on the offsets of these SNe are typically not given but are comparable to the least significant digit. We do not consider one Ca-rich SN, PTF12bho suggested to explode in the intracluster medium (since a large fraction of the stellar mass in galaxy clusters reside in the intracluster medium \cite[e.g.][and references therein]{Gon+07,Dac+08}; the existence on one such SN in our sample could be expected). For the majority of the SNe, the host galaxy is the closest galaxy and can be well identified. In two cases (PTF11kmb and PTF11bij) we followed the suggested host and offset identification in the discovery papers, although several galaxies were found nearby, and the exact identification could be debated. These might even relate to the intracluster regions where they were found. Indeed, $\sim15-20\%$ of type Ia SNe in clusters are found in the intracluster medium far from any host galaxy \citep{San+11}.  

We note that various SNe with only partial similarities to faint, type Ib, Ca-rich SNe characteristics were discussed in the literature (see e.g. \citealt{De+20,Wyn+20b} for recent overviews, and references therein); these are not considered in our sample of bona-fide Ca-rich SNe). 

\begin{table}

\centering
\caption{\label{tbl:SNe-off}
Sample of Ca-rich SNe used in this work.}
\centering
\resizebox{0.35\textwidth}{!}{
\begin{threeparttable}	\begin{tabular}{lcc}\hline	
Supernova & offset (kpc) & Galaxy type\tabularnewline
\hline 
2000ds  & 3.77  & Early\tabularnewline
2001co  & 7.07  & Late\tabularnewline
2003H  & 8.73  & Late\tabularnewline
2003dg  & 1.66  & Late\tabularnewline
2003dr  & 2.65  & Late\tabularnewline
2005E  & 24.27  & Early\tabularnewline
2005cz  & 2.12  & Early\tabularnewline
2007ke  & 16.71  & Early\tabularnewline
2010et & 37.64  & Early\tabularnewline
2012hn & 6.73  & Early\tabularnewline
PTF11bij & 34.42  & Early\tabularnewline
PTF11kmb & 150.05  & Early\tabularnewline
2016hgs & 5.9 & Late\tabularnewline
2018ckd & 19.08 & Early\tabularnewline
2018lqo & 15.46 & Early\tabularnewline
2018lqu & 26.70 & Early\tabularnewline
2018gwo & 8.56 & Early\tabularnewline
2018kjy & 6.35 & Early\tabularnewline
2019ehk & 1.8 & Late\tabularnewline
2019hty & 8.73 & Early\tabularnewline
2019pxu & 17.56 & Late\tabularnewline

\hline
\end{tabular}
\end{threeparttable}
}
\end{table} 

\subsection{Short Gamma Ray Bursts sample}
\label{sec:GRBoffset}
The offsets of sGRBs have been extensively studied by various authors. We have compiled an up-to-date list of sGRBs with published offsets and galaxy types from \citep{Fong2017,Gompertz2020,Paterson2020,O'Connor2020}. We limit the sample to those bursts with well determined offsets (i.e. error in offset determination of $\lesssim 50\%$ and excluding bursts with ambiguity in the host galaxy determination). 
Our sample consists of 8 sGRBs in early type galaxies and 15 sGRBs in late type galaxies and  is listed in table \ref{tbl:GRBoff}.

In figure \ref{fig:offsets}
we plot the cumulative offset distribution of sGRBs in early and late type galaxies (in contrast with the Ca-rich sample, some of the error-bars on the offsets are not negligible, and we therefore also show the cumulative distribution which include the error-bars in Fig.  \ref{fig:SGRBoffsets}; the overall distribution is, however, is qualitatively the same. Large offsets occur predominantly in sGRBs associated with early type galaxies. Indeed, already with the limited sample size available at the present time, the offset distributions in early and late type galaxies can be ruled out as being drawn from the same population at a level of $90\%$ confidence. This is inconsistent with the hypothesis that the offsets are dominated by strong kicks, in which case, the observed offsets (and especially for the large offset systems) will become independent of the galaxy type. 

The bottom panel of figure \ref{fig:SGRBoffsets} shows the same distributions, but now with the offsets normalized to the hosts' effective stellar radii. With this normalization, the two distributions become statistically consistent with each other, supporting the idea that the spatial extent of star formation is the main component controlling the observed offsets. We caution however that although the host normalized offset is a big step towards accounting for the galaxies' underlying stellar mass distributions, it is not in itself complete, as, especially at larger offsets, the underling stellar mass can only be revealed by ultra deep / stacked images, as discussed above.

\begin{table}
\caption{Sample of sGRBs used in this work.}
\centering
\resizebox{0.45\textwidth}{!}{
			\begin{threeparttable}	\begin{tabular}{ccccc}\hline	
					GRB &  Type & $\log_{10}(M_*/M_{\odot})$ & offset (kpc) & norm. offset \\ \hline 
					050509B & early & $11.08\pm0.03$ & $63.7\pm 12.2$ & $3.04\pm 0.58$ \vspace{0.2cm}\\ 
					050709 & late & $8.66\pm0.07$ & $3.64\pm 0.027$ & $1.75\pm 0.01$ \vspace{0.2cm}\\ 
					050724 & early & $10.64\pm0.05$ & $2.63\pm 0.079$ & $0.49\pm 0.01$ \vspace{0.2cm}\\ 
					051221A & late & $8.61\pm0.64$ & $2.18\pm 0.19$ & $0.84\pm 0.07$ \vspace{0.2cm}\\ 
					060502B & early & $\sim11.8$ & $73\pm 19$ & - \vspace{0.2cm}\\ 
					061006 & late & $10.43\pm0.23$ & $1.3\pm 0.24$ & $0.4\pm 0.07$ \vspace{0.2cm}\\ 
					070429B & late & $\sim 10.4$ & $4.7$ & - \vspace{0.2cm}\\ 
					070714B & late & $\sim 9.4$ & $12.21\pm 0.87$ & $4.56\pm 0.33$ \vspace{0.2cm}\\ 
					070724A & late & $\sim 10.1$ & $5.46\pm 0.14$ & $1.5\pm 0.04$ \vspace{0.2cm}\\ 
					070809 & early & $\sim 11.4$ & $33.22\pm 2.71$ & $9.25\pm 0.75$ \vspace{0.2cm}\\ 
					071227 & late & $\sim 10.4$ & $15.5\pm 0.24$ & $3.28\pm 0.05$ \vspace{0.2cm}\\
					080905A & late & $10.3\pm 0.3$ & $17.96\pm 0.19$ & $10.36\pm 0.1$ \vspace{0.2cm}\\
					090426 & late & - & $0.45\pm 0.25$ & $0.29\pm 0.14$ \vspace{0.2cm}\\
					090510 & late & $\sim 9.7$ & $10.37\pm 2.89$ & $1.99\pm 0.39$ \vspace{0.2cm}\\
					090515 & early & $\sim 11.2$ & $75.03\pm 0.15$ & $15.53\pm 0.03$ \vspace{0.2cm}\\
					100117A & early & $\sim 10.3$ & $1.32\pm 0.33$ & $0.57\pm 0.13$ \vspace{0.2cm}\\
					111117A & late & $9.9\pm 0.2$ & $8.5\pm 1.7$ & - \vspace{0.2cm}\\
					120804A & late & $\sim 10.8$ & $2.2\pm 1.2$ & - \vspace{0.2cm}\\
					130603B & late & $\sim 9.7$ & $5.21\pm 0.17$ & $1.05\pm0.04$ \vspace{0.2cm}\\
					150101B & early & $10.85\pm 0.12$ & $7.35\pm 0.07$ & $0.77\pm0.02$ \vspace{0.2cm}\\
					170817A & early & $10.65\pm 0.03$ & $2.12\pm 0.001$ & $0.64\pm0.03$ \vspace{0.2cm}\\
					181123B & late & $10.24\pm 0.16$ & $5.08\pm 1.38$ & - \vspace{0.2cm}\\
					200522A & late & $9.44\pm 0.02$ & $1.07\pm 0.27$ & $4\pm0.1$ \vspace{0.2cm}\\
					\hline
					\label{tbl:GRBoff}
				\end{tabular}
			\end{threeparttable}
}
\end{table}

\section{Discussion}
\label{sec:discussion}
\subsection{The progenitors of Ca-rich supernovae}
Ca-rich SNe are faint (>$\sim$16(16.5) B(R) Mag) type Ib SNe, with strong Ca to O line ratio \citep{Per+10}. The prototype SN 2005E was found in the halo of an early type, S0 galaxy at a large offset ($\sim23$ kpc) from its host galaxy center. At the time, all type Ib SNe were thought to arise from the core-collapse of a massive star. We therefore initially thought that SN 2005E progenitor was a massive star that formed in the galaxy nucleus (where gas and star formation might exist even in an early type galaxy), and was later ejected as a hypervelocity star following an interaction with the central massive black hole. It then propagated in the galaxy and exploded in its observed remote location in the galaxy halo. Although potentially possible, our analysis showed that the likelihood of observing an event from such a scenario was highly unlikely \citep{Per+10}. 

After collecting and characterizing a sample of such SNe, we found many of them to explode in old stellar environments, and some at large offsets similar to SN 2005E.  Together with our finding of inferred low ejecta-mass, low energy, low velocities and low $^{56}{\rm Ni}$ mass, as well as potentially large abundances of intermediate elements (and the lack of any star-forming regions close-by) we concluded that Ca-rich SNe likely constitute a novel type of SN explosion arising from a thermonuclear He-rich explosion on a relatively old white dwarf \citep{Per+10}. We therefore suggested that this scenario well explains the properties of these SNe, and the existence of such type Ib SNe in old stellar environments such as early type galaxies and galaxy halos \citep{Per+10,Wal+11,Per+11,Kasl+12,Lym+13,Per+14}.

The majority of Ca-rich SNe were found in old stellar populations in early type galaxies \cite[e.g.]{Per+10,De+20}
it was therefore expected that the locations and offsets of such SNe in their host galaxies would follow the old stellar populations in these galaxies. However, many of the SNe both in our original \citep{Per+10} sample and additional Ca-rich SNe identified later on \citep{Kasl+12,De+20} were found at large offsets from their host galaxies (see Table \ref{tbl:SNe-off} and Fig. \ref{fig:offsets}), where very little, if any, stellar population was thought to exist.

As we first briefly suggested in \cite{Per+14} and now explore in detail in this study, the large offsets are in fact, a natural outcome from the existence of extended stellar halos of (mostly) early type galaxies \citep{dso+14,Hua+18}, as discussed in section \ref{sec:halos}.  

As can be seen in Fig. \ref{fig:offsets}, The offsets distribution in early type and late galaxies show a different behaviour, with most of the SNe in late galaxies residing in the central (<10 kpc) parts, while most of the SNe in early-type galaxies residing in the halo (>10 kpc), nicely consistent with the stellar mass fractions for late and early type galaxies, respectively, as found in observations  \citep{dso+14,Hua+18}. 
In fact, in retrospect, the large offset distribution of these SNe, and its difference in early vs. late type galaxies (first shown by us here in Fig. \ref{fig:offsets}) anticipated these findings.   

The perplexing findings of large offsets instigated several studies suggesting that the remote locations of such transients could be explained either by large velocity kicks given to their progenitors either because their progenitors were NSs receiving a natal-kick at birth.

\cite{Yua+13} and later \cite{Sel+15} and \cite{She+19} suggested that the extended distribution of the locations of Ca-rich is consistent
with the distribution of globular clusters distribution. However, photometric searches for globular clusters close to the positions of
known Ca-rich transients have generally been unsuccessful \citep{Lym+14}. Furthermore,  the relatively high inferred rate of Ca-rich SNe (5-15$\%$ of the Ia SNe rate; \citealt{Per+10,De+20}) could be difficult to explain with a globular cluster origin, given the low mass fraction in globular clusters. 

Others suggested a large velocity kick is imparted to the progenitors of Ca-rich SNe \citep{Lym+14,Fol+15,Lym+16}. Such scenarios are difficult to reconcile with the suggested WD merger/He-accretion origins of Ca-rich SNe \citep{Per+10,She+10,Wal+11,Des+15,Per+19}. They are also inconsistent with the expected rates and delay time distribution NS-WD mergers origins \citep{Too+18,Zen+19}, suggested as progenitors for Ca-rich SNe \citep{Met+12}, and with the ejection rates of hypervlocity stars by massive black holes \citep{Per+10}. Furthermore, as we showed here, the offsets in early type galaxies appear to extend farther than in late type galaxies, raising an extra challenge to a kick origin which relies on the micro-physics of the progenitor rather than on the galaxy macro-physics. 

The results shown here therefore reconcile the difficulties and inconsistencies inferred form the large sample of Ca-rich SNe with large offsets.
We find the large \emph{apparent} offsets of Ca-rich SNe from the underlying stellar distribution is just an artifact of the low-surface stellar brightness in galaxy halos, which is difficult to resolve with regular imaging, and requires the use of stacked or deep imaging. We conclude that Ca-rich SNe mostly arise from an old stellar population, explaining the large fraction of early type galaxies hosts. The high frequency of early type hosts and the large extended stellar halos of such galaxies explain the large offset distribution among Ca-rich SNe. 

We note that in early type galaxies the fraction of Ca-rich SNe in the central 2-3 kpc appears to be lower than that of sGRBs in such galaxies. This might only reflect the low-statistics, but it is important to note that one should expect lower detection rates in the denser inner regions of galaxies, especially for very subluminous SNe such as Ca-rich SNe, as already suggested by us \citep{Per+14}, i.e. we expect significant Shaw's selection bias \citep{Sha79} for such SNe. 

Finally, early type galaxies have higher metallicities than late-type galaxies \citep[e.g]{Li+18}. We therefore raise the possibility that Ca-rich SNe progenitors originate from \emph{high} metallicity environments, rather than low-metallicities as suggested by us and others in the past \citep{Yua+13,Per+14,She+19}. However, detailed study of the host metallicities for Ca-rich SNe has yet to be done.  

\subsection{The kick velocity of sGRB progenitors}
The binary neutron star (BNS) merger origin for sGRBs has been suggested decades ago \citep{Goodman1986,Eichler1989}. It received a conclusive proof with the detection of GRB 170817, a GRB accompanying a GW trigger from a BNS merger \citep{Abb+17}. Since BNSs receive kicks when the second star in the binary collapses to a neutron star, they could potentially travel significantly before merging and producing a sGRB. An estimation of these kicks, merger delay times and offset distances can be obtained from the Galactic population of BNSs.

Shortly after the discovery of the double pulsar J0737-3039  \citep{Burgay2003}, the orbital parameters of this system led  \cite{Piran2004,Piran2005} to suggest that the younger pulsar in the system, pulsar B, must have formed with very small mass ejection and with virtually no kick velocity. The system had a very small eccentricity and it was located in the Galactic plane. The small eccentricity suggested a small mass ejection, unless a strongly fine-tuned kick was given to the system. However, the latter would have led to a large CM velocity and the location of the system in the Galactic plane indicated that at least the vertical (out of the plane) component of the  velocity of the system was low. The very small proper motion predicted by \cite{Piran2004,Piran2005} was confirmed within a year by pulsar timing observations \citep{Kramer2006}. \cite{DallOsso2014} refined the original estimates, using the observed proper motion and other parameters of the system. They confirmed the earlier expectations and have shown that in this system, the older pulsar (A) also most likely formed in this way. 

Later on, \cite{Beniamini2016} have shown that out of 10 BNS systems observed in the Galaxy that don't reside in globular clusters, between 6 and 7 must have been formed by collapses involving the release of a small amount of mass $\Delta M\lesssim 0.5 M_{\odot}$ and imparting only a very weak kick $v_k\lesssim 30 \mbox{km/sec}$, corresponding to changes in the center of mass velocity of the BNS of $\Delta v_{\rm cm}\lesssim 10\mbox{km s}^{-1}$. The other BNS systems require more typical core-collapses with an ejected mass of one to a few solar masses and kicks of a few hundred km/sec. Since the publication of \cite{Beniamini2016}, four new BNS systems have been reported: J1913+1102, J1757-1854, J1411-2551, J1946+2052 \citep{Lazarus2016,Martinez2017,Cameron2018,Stovall2018}. Three out of these four have small eccentricities $e=0.06,0.09,0.17$, suggesting a weak collapse origin and only one has $e=0.6$. This ratio is in a nice agreement with the earlier expectations and suggests that the result is not dominated by a statistical fluctuation. Furthermore, these conclusions were independently reproduced by \cite{Tauris2017}.
Overall, approximately $2/3$ of BNS systems observed so far require weak collapses with small kicks to account for their observed orbits.

\begin{figure}
\centering
\includegraphics[width = 0.4\textwidth]{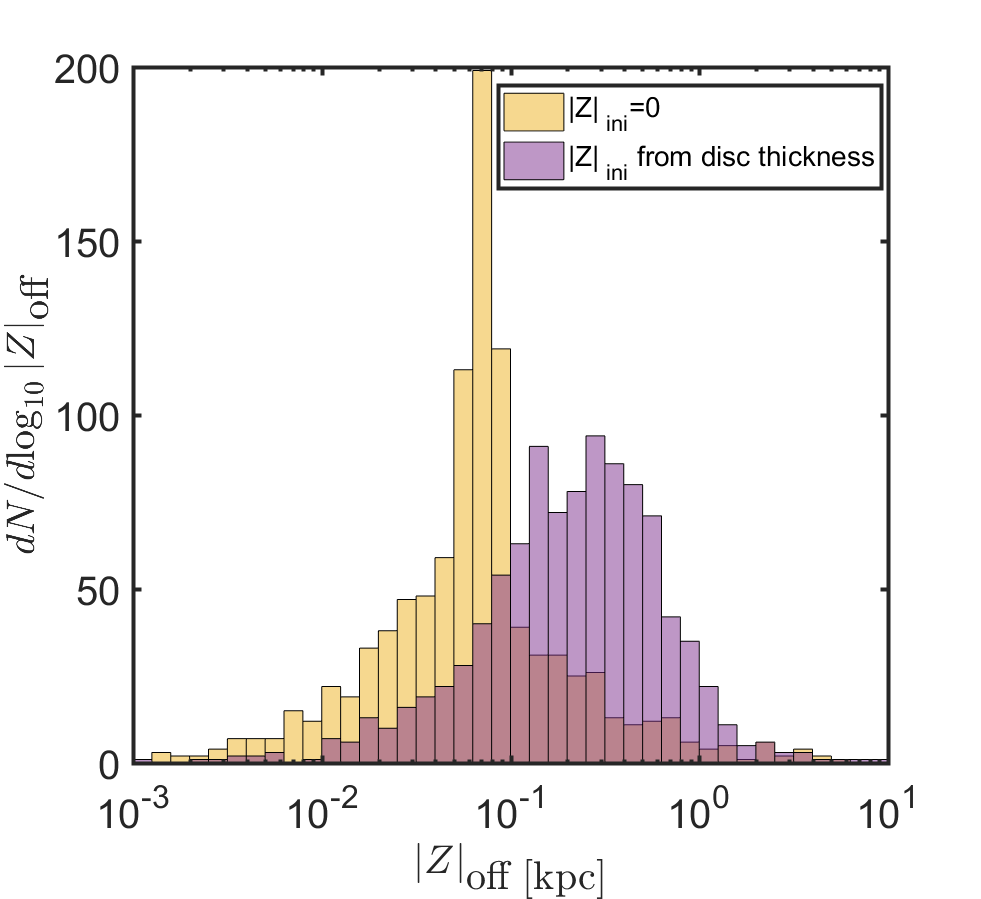}
\caption{Results of a Monte Carlo simulation calculating the vertical offsets, $|z|_{\rm off}$, of BNS systems from the Galactic plane at the time of merger (taking into account constraints on kicks and time to merger, consistent with constraints from the Galactic population, see \S \ref{sec:zMW} for details). We show results with the initial vertical offsets, $|z|_{\rm ini}$, either in the Galactic plane (yellow) or following the disc thickness (purple). The results demonstrate that the offsets induced by kicks alone are typically $\lesssim 0.1$kpc, and suggest that the vertical offset distribution at merger is dominated by the initial vertical spread.}
\label{fig:heightsims}
\end{figure}

With the small center of mass kicks mentioned above, BNS systems cannot easily break away from their host galaxy's potential and travel to large offsets. An illustrative example is given by the Milky Way. Assuming BNS systems are born close to the Galactic disc, and applying the small kicks mentioned above, will induce vertical oscillations above the Galactic plane, with a maximal length scale of $z_{\rm typ}\sim \Delta v_{\rm cm} P_z/2\pi\sim 0.08 \mbox{ kpc}$, where $P_z\sim 50 \mbox{ Myr}$ is the time-scale for $z$ oscillations in the potential of the Galaxy. This is more than two orders of magnitude lower than the offsets measured for some sGRBs in galaxies with similar stellar mass to that of the Milky Way (see table \ref{tbl:GRBoff}).
Therefore, if BNSs are born close enough to the Galactic plane, that their motion is dominated by the Galactic potential, the expected distribution of kicks does not naturally account for large off-sets. To demonstrate the latter point we plot in figure \ref{fig:heightsims} the vertical offset distribution of BNS systems at the time of merger, as resulting from a Monte Carlo simulation, in which the birth conditions of the binaries are informed by constraints from the Galactic BNS population and the systems' motion in the Galaxy's potential is then calculated with galpy \footnote{\url{http://github.com/jobovy/galpy}} \citep{Bovy2015} using the \texttt{MWPotential2014} Galactic potential (full details of the Monte Carlo calculation are given in appendix \ref{sec:zMW}). The results re-affirm the OoM estimate for $z_{\rm typ}$ above, and the tension with sGRB offsets. 

Of course, if BNS systems can be formed far away from the centers of their host galaxies, large off-sets can be naturally explained with no need for invoking kicks \footnote{We note, however, that at large offsets, the galactic potentials can become sufficiently weak such that the BNS can travel unimpeded, even with a very small change to its center of mass. The BNS will then cover a distance $\approx t_{\rm mrg}\Delta v_{\rm cm}$ before merging (where $t_{\rm mrg}$ is the delay time between BNS formation and merger). Recently, \citep{BP2019} have shown that the lifetime of the pulsars in Galactic BNS systems are sufficiently short, that the delay times may be directly inferred from the currently observed time until merger in those systems. The result is a rather steep delay time distribution in which at least $40-60\%$ ($10-20\%$) of systems were formed with delays shorter than 1 Gyr (100 Myr). Therefore, even in this case BNS systems are unlikely to move very far (compared to their offset at birth) before eventually merging.}. In such a situation, the offsets of BNS mergers should correlate with the spatial extent of star formation in their host galaxies. This too is demonstrated in the results of our Monte Carlo calculation for the Milky Way presented in figure \ref{fig:heightsims}. This also naturally leads to a systematic difference between the offsets of BNS mergers in early vs late time galaxies as seen in observations (see section \ref{sec:offsets}).

\section{Summary}
Ca-rich SNe and sGRBs are two groups of transients, of which members show an extended spatial distributions in their host galaxies. In particular, a large fraction of these transients are found at large (>10 kpc) from the nuclei of their galaxy hosts, and apparently far from any observed underlying stellar population, which could host their progenitors. These perplexing findings motivated various solutions. These include models in which the progenitors of such transients received large velocity kicks allowing them to migrate to remote positions where they exploded; or that the progenitors originated in globular clusters, the distribution of which extend far into the halo. However, such solutions are difficult to reconcile with other properties of the transients and their suggested progenitors, as we discussed above. 

In this study we analyzed the distributions of the distance offsets of Ca-rich supernovae and sGRBs from their host galaxies. 
We pointed out that stacked-images and deep-imaging of early and late type galaxies show different stellar mass distribution, and that early type galaxies have very extended stellar mass, with a large fraction, typically the majority of the stellar mass resides in the halo beyond 10 kpc. Therefore, in contrast to previous studies, we divided the offsets distributions between early and late type galaxies. We showed that the offsets in early type galaxies extend farther than that in late-type galaxies. Moreover, the fractions of transients in the halo (>10 kpc) vs. the central parts (<10 kpc) are consistent with the observed distinctly different halo to central parts stellar mass  ratios in early type and late type galaxies \citep{dso+14,Hua+18}.

We also studied the expected offset distribution for progenitors that receive velocity kicks and show that no large kicks are required in order to  explain the observed offsets.  

We conclude that the progenitors of both Ca-rich SNe and sGRBs do not require large velocity kicks, nor do their progenitor need to form in globular clusters. Rather, their offset distributions are generally consistent with the underlying stellar mass distribution of their host galaxies. The appearance of large offsets is then consistent with old stellar progenitors for significant fractions of these transients. 

Finally, we note that since stacked/ultra-deep imaging show that early-type galaxies are more extended than late-type galaxies, any study of other transients' offset distribution, such as FRBs should account for the host galaxy-type.  In fact, the recent finding of FRBs at large offsets \citep{Ravi2019}, does not require their progenitors to have had natal kicks, and we expect FRBs with large offsets to be mostly identified in early type galaxies.

\label{sec:summary}
\section*{Data availability}
The simulations underlying this article will be shared on reasonable request to the corresponding author.

\section*{Acknowledgements}
HBP acknowledges support for this project from the European Union's Horizon 2020 research and innovation program under grant agreement No 865932-ERC-SNeX. The research of PB was funded by the Gordon and Betty Moore Foundation.

\bibliographystyle{mnras}

\appendix

\section{Host normalized sGRB offsets}
\begin{figure}
\centering
\includegraphics[width = 0.4\textwidth]{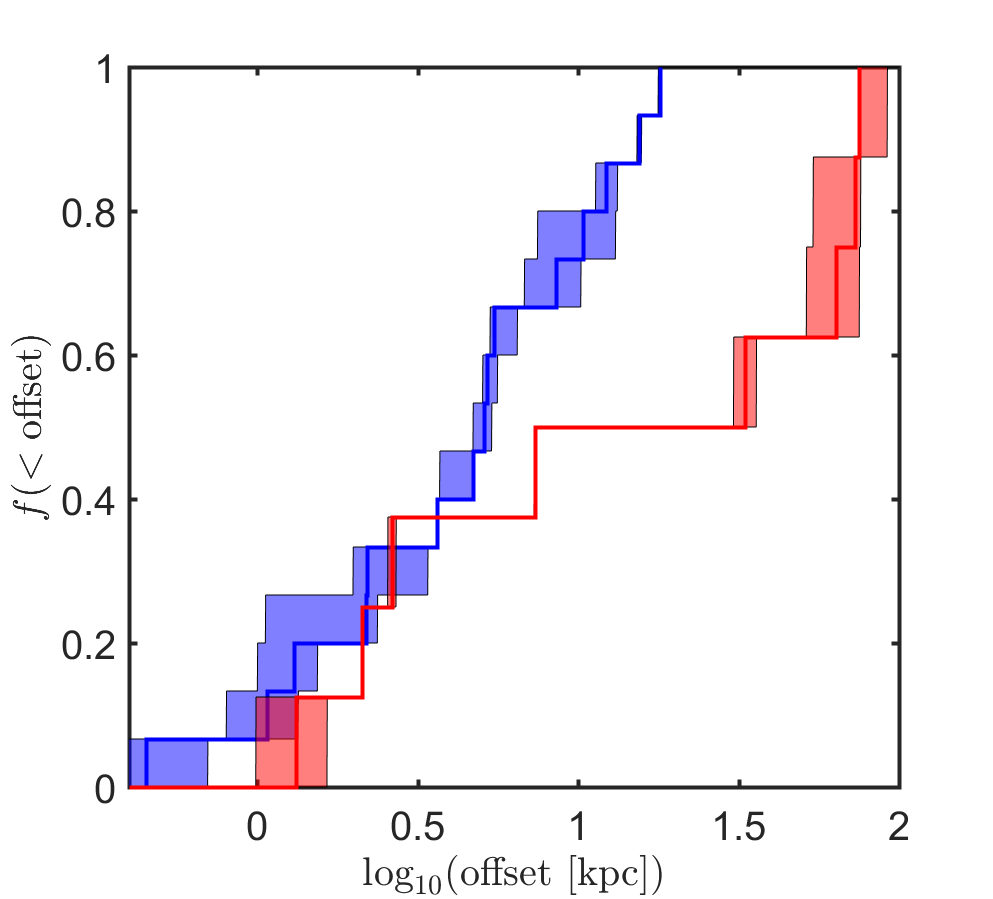}
\includegraphics[width = 0.4\textwidth]{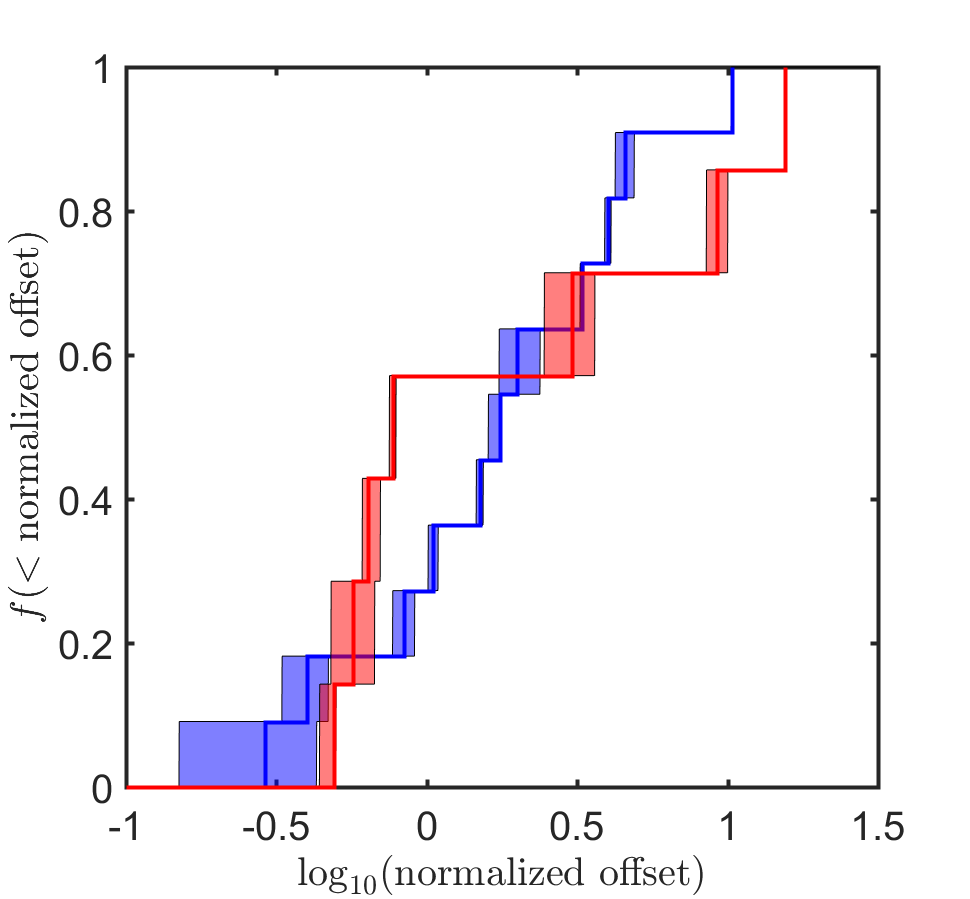}
\caption{Top: cumulative distribution of sGRB offsets. Results are separated to early-type galaxies (blue) and late-type (red). Bottom: Same as above, but for the offsets normalized by the respective effective stellar radii of each galaxy.}
\label{fig:SGRBoffsets}
\end{figure}

We plot in Fig. \ref{fig:SGRBoffsets} the cumulative distribution of sGRB offsets separating different galaxy types and considering also the host normalized offsets (see \S \ref{sec:GRBoffset} for details).

\section{Monte Carlo simulation of BNS merger vertical offsets}
\label{sec:zMW}
We apply a Monte Carlo calculation to estimate the distribution of vertical offsets of BNS mergers above the Galactic plane. We begin by drawing the birth locations of the binaries. The systems are assumed to either (i) be born in the Galactic plane (this allows to isolate the effect of kicks from the initial height distribution) or (ii) be drawn from a distribution $d{\rm P}/dz\propto \exp(-r/h_z)$ with $h_z=0.3$\,kpc (matching the adopted \texttt{MWPotential2014} Galactic potential \citep{Bovy2015}). In either case, their Galactic radius is drawn from a distribution $d{\rm P}/dr\propto \exp(-r/h_r)$ with $h_r=2.6$\,kpc (again matching \texttt{MWPotential2014}). The systems are initially rotating in the Galactic with their tangential velocity set by the \texttt{MWPotential2014} rotation curve. Next, we randomize the kicks and mass ejection in the superonova leading to the the birth of the second NS in the binary. Following \cite{Beniamini2016} we take log-normal distributions with median values of $\Delta M_0, v_{\rm k,0}$ respectively and with widths $\sigma_{\Delta M}/\Delta M_0=\sigma_{v_{\rm k,0}}/v_{\rm k,0}=0.5$. As per \cite{Beniamini2016}, 2/3 of the systems surviving the supernova (and merging within less than a Hubble time) have weak collapses and small mass ejection, $\Delta M_0=0.05M_{\odot},v_{\rm k,0}=5\mbox{km s}^{-1}$ while the other 1/3 have values more typical of standard core-collapse supernovae, $\Delta M_0=1M_{\odot},v_{\rm k,0}=160\mbox{km s}^{-1}$. We also include an additional amount of ejecta, $\Delta M_{\nu}=0.1M_{\odot}$ due to neutrino emission \citep{BHP2016}. The direction of the kicks is assumed to be uncorrelated with the orientation of the binary's orbit. For the latter, we also assume the orbit to be circular before the supernova takes place with the initial separation of the systems following a PL distribution $dN/da\propto a^{-2.2}$ above $a_{\rm min}=7.5\times 10^{10}$\,cm. The latter distribution is consistent with the observed Galactic binaries and reproduces their observed delay time distribution between second NS formation, and BNS merger \citep{BP2019}. With these ingredients in place, we self-consistently calculate the resulting orbit in each system, the time between its formation and merger and its center of mass velocity (see \citealt{BHP2016} for more details). We then propagate the motion of the systems in the Galactic potential using galpy \citep{Bovy2015} until the time of merger. We repeat this process $10^3$ times and record the distribution of heights of merging binaries above the Galactic plane.

\bsp	
\label{lastpage}
\end{document}